\documentclass[conference]{IEEEtran}
\IEEEoverridecommandlockouts

\usepackage{cite}
\usepackage{amsmath,amssymb,amsfonts}
\usepackage{algorithmic}
\usepackage{graphicx}
\usepackage{textcomp}
\usepackage{url}
\usepackage{hyperref}

\usepackage{bm}
\makeatletter
\AtBeginDocument{\DeclareMathVersion{bold}
\SetSymbolFont{operators}{bold}{T1}{times}{b}{n}
\SetMathAlphabet{\mathrm}{bold}{T1}{times}{b}{n}
\SetMathAlphabet{\mathit}{bold}{T1}{times}{b}{it}
\SetMathAlphabet{\mathbf}{bold}{T1}{times}{b}{n}
\SetMathAlphabet{\mathtt}{bold}{OT1}{pcr}{b}{n}
\SetSymbolFont{symbols}{bold}{OMS}{cmsy}{b}{n}
\renewcommand\boldmath{\@nomath\boldmath\mathversion{bold}}}
\makeatother

\def\BibTeX{{\rm B\kern-.05em{\sc i\kern-.025em b}\kern-.08em
    T\kern-.1667em\lower.7ex\hbox{E}\kern-.125emX}}

\newcommand{\Zq}{\mathbb{Z}_q}

\newcommand{\B}{{\mathcal B}}

\newcommand{\Sig}{{\sf SIG}}
\newcommand{\sigvk}{{\sf vk}_{\sf sig}}
\newcommand{\sigsigk}{{\sf sigk}_{\sf sig}}
\newcommand{\SigKeyGen}{{\sf Sig.KeyGen}}
\newcommand{\SigVerify}{{\sf Sig.Verify}}
\newcommand{\SigSign}{{\sf Sig.Sign}}

\newcommand{\ECDSA}{{\sf ECDSA}}
\newcommand{\addr}{{\sf addr}}

\newcommand{\IBS}{{\sf IBS}}

\newcommand{\ibssigk}{{\sf sigk}_{\sf ibs}}
\newcommand{\IBSSetup}{{\sf IBS.Setup}}
\newcommand{\IBSKeyDer}{{\sf IBS.KeyDer}}
\newcommand{\IBSVerify}{{\sf IBS.Verify}}
\newcommand{\IBSSign}{{\sf IBS.Sign}}
\newcommand{\mpk}{{\sf mpk}}
\newcommand{\msk}{{\sf msk}}
\newcommand{\cert}{{\sf cert}}
\newcommand{\ibssig}{\sigma_{\sf ibs}}

\RequirePackage{xcolor}

\usepackage{listings}
\definecolor{verylightgray}{rgb}{.97,.97,.97}

\lstdefinelanguage{Solidity}{
	keywords=[1]{anonymous, assembly, assert, balance, break, call, callcode, case, catch, class, constant, continue, constructor, contract, debugger, default, delegatecall, delete, do, else, emit, event, experimental, export, external, false, finally, for, function, gas, if, implements, import, in, indexed, instanceof, interface, internal, is, length, library, log0, log1, log2, log3, log4, memory, modifier, new, payable, pragma, private, protected, public, pure, push, require, return, returns, revert, selfdestruct, send, solidity, storage, struct, suicide, super, switch, then, this, throw, transfer, true, try, typeof, using, value, view, while, with, addmod, ecrecover, keccak256, mulmod, ripemd160, sha256, sha3}, % generic keywords including crypto operations
	keywordstyle=[1]\color{blue}\bfseries,
	keywords=[2]{address, bool, byte, bytes, bytes1, bytes2, bytes3, bytes4, bytes5, bytes6, bytes7, bytes8, bytes9, bytes10, bytes11, bytes12, bytes13, bytes14, bytes15, bytes16, bytes17, bytes18, bytes19, bytes20, bytes21, bytes22, bytes23, bytes24, bytes25, bytes26, bytes27, bytes28, bytes29, bytes30, bytes31, bytes32, enum, int, int8, int16, int24, int32, int40, int48, int56, int64, int72, int80, int88, int96, int104, int112, int120, int128, int136, int144, int152, int160, int168, int176, int184, int192, int200, int208, int216, int224, int232, int240, int248, int256, mapping, string, uint, uint8, uint16, uint24, uint32, uint40, uint48, uint56, uint64, uint72, uint80, uint88, uint96, uint104, uint112, uint120, uint128, uint136, uint144, uint152, uint160, uint168, uint176, uint184, uint192, uint200, uint208, uint216, uint224, uint232, uint240, uint248, uint256, var, void, ether, finney, szabo, wei, days, hours, minutes, seconds, weeks, years},	% types; money and time units
	keywordstyle=[2]\color{teal}\bfseries,
	keywords=[3]{block, blockhash, coinbase, difficulty, gaslimit, number, timestamp, msg, data, gas, sender, sig, value, now, tx, gasprice, origin},	% environment variables
	keywordstyle=[3]\color{violet}\bfseries,
	identifierstyle=\color{black},
	sensitive=false,
	comment=[l]{//},
	morecomment=[s]{/*}{*/},
	commentstyle=\color{gray}\ttfamily,
	stringstyle=\color{red}\ttfamily,
	morestring=[b]',
	morestring=[b]"
}

\begin{document}

\title{On the Consideration of Vanity Address Generation via Identity-Based Signatures}

\author{
\IEEEauthorblockN{Shogo Murasaki\thanks{This work was done when he was in Kanazawa University. He is currently a master student at Institute of Science Tokyo.}}
\IEEEauthorblockA{\textit{Kanazawa University, Japan}}\\
\and
\IEEEauthorblockN{Kazumasa Omote}
\IEEEauthorblockA{\textit{University of Tsukuba, Japan}}\\
\and
\IEEEauthorblockN{Keita Emura\thanks{Corresponding Author: k-emura@se.kanazawa-u.ac.jp}}
\IEEEauthorblockA{\textit{Kanazawa University, Japan}}
\IEEEauthorblockA{\textit{AIST, Japan}}\\
}

\maketitle

\begin{abstract}
An address is indicated as an identifier of the user on the blockchain, and is defined by a hash value of the ECDSA verification key. A vanity address is an address that embeds custom characters such as a name. To generate a vanity address, a classical try-and-error method is employed, and thus the number of characters to be embedded is limited. 
In this paper, we focus on the functionality of identity-based signatures (IBS) where any strings can be employed as a verification key, and explore whether IBS can be used for generating a vanity address. 
We attach importance to the fact that it is not realistic to replace ECDSA with key recovery, which is currently employed for issuing transactions in Ethereum, to an IBS scheme. Even if this replacement is possible, it is not a reasonable price for the ease of the vanity address generation. Thus, we pay attention to a generic construction of IBS from signatures, and construct an IBS scheme from ECDSA with key recovery. Though we cannot directly generate a vanity address due to the key recovery functionality of the underlying ECDSA, we can connect any string with an address due to the functionality of IBS that can give additional meaning to the address. We implement our system by Solidity, and demonstrate that the gas cost is almost same as that of the ECDSA signature verification.  
\end{abstract}

\begin{IEEEkeywords}
Blockchain, Vanity address, Identity-based signatures, ECDSA with key recovery 
\end{IEEEkeywords}

\section{Introduction}
\IEEEPARstart{A} user signs a transaction using own secret ECDSA signing key when the user issues the transaction in Ethereum, where ECDSA stands for elliptic curve digital signature algorithm. 
An address is indicated as an identifier of the user on the blockchain, and is defined by a hash value of the ECDSA verification key. Precisely, Ethereum Yellow Paper~\cite{ETHYellow} writes that \lq\lq A Ethereum address is defined as the rightmost 160-bits of the Keccak-256 hash of the corresponding ECDSA public key". 
Here, an address is looked as a random value because the underlying verification key is generated by a secret signing key that is chosen at random. It is not trivial to check whether an address is valid (meaning that the address is managed by an expected user).  

\medskip
\noindent\textbf{Vanity Address}. A vanity address is an address that embeds custom characters such as a name. It is not Ethereum-specific. For example, Edward Snowden publishes a public key of Nostr \lq\lq npub1sn0wdenkukaku0d9df....".%
\footnote{\url{https://x.com/Snowden/status/1620790688886718466}} 
Here, sn0wden (o is replaced to 0) is embedded, and it can be seen as a vanity address. 
An article (Top 5 Bitcoin Vanity Addresses~\cite{Top5}) reports an address embedding the currently longest character \lq\lq Embarassable" (actual value is EMBARraSS) and a palindrome address \lq\lq 1234m....U4321". From these examples, it seems that the motivation to embed some meaningful string/character in an address, that is essentially a random value, is relatively popular. 
We also introduce vanity URLs. The article~\cite{VanityURLs} introduces vanity URLs as follows: 

\begin{quote}
\lq\lq \emph{custom-made, easy-to-remember, and they pack a punch when it comes to branding}". 
\end{quote}

\noindent 
Moreover, it insists that 

\begin{quote}
\lq\lq \emph{But why should you care about vanity URLs? Well, they’re essential for your brand. They make your web address more user-friendly, more memorable, and they can significantly boost your SEO. They target your audience (buyers) specifically, enhancing the user experience all around. So, if you’re serious about taking your branding to the next level, it’s time to get familiar with vanity URLs.}
\end{quote}

\noindent
Though this is an article about vanity URLs, it also well explains the effectiveness of vanity address as well .

\medskip
\noindent\textbf{How to Generate a Vanity Address at Present}. 
A vanity address is generated by a classical try-and-error method as follows. 

\begin{enumerate}

\item Choose a secret signing key at random. 
\item Generate the verification key.
\item Check whether its hash value contains the expected character, and repeat this cycle until the desired result is obtained. 

\end{enumerate}

\medskip
\noindent\textbf{Issues in the Current Generation Method of Vanity Addresses}. Here, we discuss issues of the above classical try-and-error method in terms of computational costs and the number of characters to be embedded.  
To generate a verification key, algebraic operations (additions over an elliptic curve in the case of ECDSA) are required. These operations are quite inefficient compared to computations of a hash function. Moreover, a secret signing key needs to be chosen at random (to prevent the guessing attack) and thus generating a vanity address will not be completed in a realistic amount of time when a relatively long character is considered to be embedded. 
An article~\cite{BitcoinVanity} introduces that a Bitcoin vanity address can be generated by one hour when 5 characters are considered to be embedded, but it requires more than three months if 7 characters are considered to be embedded. Concretely, we tried to generate a Ethereum vanity address using VANITY-ETH~\footnote{ETH vanity address generator: \url{https://vanity-eth.tk/}}. For 7 characters, it displayed that \lq\lq 50\% probability: 5 hours, 11 minutes", for 8 characters, it displayed that \lq\lq 50\% probability: 3 days, 7 hours", and for 9 characters, it displayed that \lq\lq 50\% probability: 1 month, 3 weeks". On the other hands, we can generate a vanity address in seconds when 3 characters are indicated. 
We remark that, in addition to the number of characters to be embedded, it is restricted that only 0-9 and A-F are possible in Ethereum since each address is expressed by hexadecimal numbers. 

\medskip
\noindent\textbf{Real Incident}. 
To reduce the computational cost,  we should not use a relatively small random number to efficiently generate a vanity address. 
Actually, a vulnerability of a tool for vanity address generation, Profanity, has been reported where a 32-bit seed was set for generating a random number (See CVE-2022-40769~\cite{CVE-2022-40769}), and a cryptocurrency market maker, Wintermute, was hacked for around \$160 million in September 2022.

\medskip
\noindent\textbf{Our Motivation}. 
In view of the above situation, it seems a natural motivation to embed a character  with some meaning to an address, but the number of character is limited as a few words and vanity address generation takes an enormous amount of time. 
Though the above incident seems an implementation vulnerability, it is highly desirable to propose a method to safety and efficiently generate a vanity address. 

As the first attempt, we focus on the functionality of identity-based signatures (IBS)~\cite{Shamir84} where any strings, say identity $ID$ in the IBS context, can be employed as a verification key, and explore whether IBS can be used for generating a vanity address as follows (we will explain that the attempt fails later). 

\begin{enumerate}

\item Indicate a character to be embedded to a vanity address. 

\item Choose a 128-bit random number. 

\item Compute its hash value, and repeat this cycle until the desired result is obtained. 

\item Set the random value as $ID$ when a desired address is generated. 
\end{enumerate}

\noindent
Then, it is expected that the number of algebraic operations can be drastically reduced compared to the classical try-and error method. We remark that a key generation center (KGC) is defined that issues a secret signing key for $ID$ using the own master secret key. Then, a key escrow problem happens where the KGC also can generate signatures. 
To solve the problem, TEE (Trusted Execution Environment) could be employed (See Section~\ref{proposed}).

\medskip
\noindent\textbf{Limitation of the First Attempt}. 
Though IBS seems a promising tool to safety and efficiently generate a vanity address, we attach importance to the fact that it is not realistic to replace ECDSA with key recovery, which is currently employed for issuing transactions in Ethereum, to an IBS scheme. Even if this replacement is possible, it is not a reasonable price for the ease of the vanity address generation. 

\medskip
\noindent\textbf{Our Contribution}. 
In this paper, we pay attention to the fact that IBS can be generically constructed from any signature scheme~\cite{BellareNN04,DodisKXY03,GentryS02}, and construct an IBS scheme from ECDSA with key recovery. 
Then, we consider whether a vanity address can be generated by using the ECDSA-based IBS scheme via the above attempt. 
Counterintuitively, we cannot directly generate a vanity address. 
We stress that, although it can be seen as a negative result, clarifying this counterintuitive fact is also our important contribution. 

On the other hands, we can connect any string with an address due to the functionality of IBS that can give additional meaning to the address. 
For example, a name can be written together with an address, and the unforgeability of IBS \emph{cryptographically} guarantees that the name is connected to the address. This is a crucial difference from the case that a name is just written together with an address. 

We implement our system by Solidity, and demonstrate that the gas cost is twice compared to that of the ECDSA signature verification. 
We further pointed out that one of two ECDSA signature verification is independent to the message (transaction) in the IBS signature verification procedure, and is for verification of a certification. 
Since the certification needs to be verified only once, the actual gas cost for verifying a transaction is essentially same as that of the ECDSA signature verification.  

\medskip
\noindent\textbf{Related Work}. Baldimtsi et al.~\cite{BaldimtsiCCK22} showed that the security is preserved even if a part of the output of a hash function is previously indicated in terms of the bit security framework defined by Watanabe and Yasunaga~\cite{WatanabeY21}. 
Though the main purpose of Baldimtsi et al. is to reduce the storage cost by fixing a part of hash value, they mentioned about vanity addresses.   
To the best of our knowledge, this is the only work that considers vanity addresses from the cryptographic point of view, and they did not consider IBS in their paper.  

\section{ECDSA with Verification Key Recovery}

In the conventional signature scheme, the verification algorithm takes a verification key in addition to a signature and a message. Concretely, let $\Sig=(\SigKeyGen,\SigSign,\SigVerify)$ be a signature scheme, $(\sigvk,\sigsigk)\leftarrow\SigKeyGen(1^\lambda)$ be a pair of a verification key and a signing key (here, $\lambda\in\mathbb{N}$ is a security parameter), $\sigma\leftarrow\SigSign(\sigsigk,M)$ be a signature on a message $M$. Then, the verification algorithm is run such that $\SigVerify(\sigvk,\sigma,M)$. 
However, Ethereum employs ECDSA with key recovery where a verification key is recovered from $(sigma,M)$ and the verification algorithm does not take a verification key as input. 
Concretely, check whether the hash value of the recovered verification key is equal to an address. Precisely, an address is described as $\addr=\B_{96..255}(H(\sigvk))$ where $\B_{96..255}$ is the rightmost 160-bits of the Keccak-256 hash (See Ethereum Yellow Paper~\cite{ETHYellow}). 
This ECDSA with key recovery is employed as the underlying signature scheme to construct an IBS scheme via the generic construction~\cite{BellareNN04,DodisKXY03,GentryS02}. The selection is reasonable when the proposed system is considered to be implemented in the actual blockchain environment. 

Next, we introduce ECDSA with key recovery as follows. Let $p$ and $q$ be prime numbers, $H:\{0,1\}\rightarrow\Zq$ be a hash function, $E/\mathbb{F}_p$ be an elliptic curve with order $q$ defined over $\mathbb{F}_p$, and $G\in E(\mathbb{F}_p)$ be a base point. 
Assume that each algorithm implicitly takes $(E,G,p, q)$ as input. 
We describe a point on $E$ as $R=(R_x,R_y)$. 
Here, if $R=(R_x,R_y)$ is a point on $E$, then $-R=(R_x,-R_y)$ is a point on $E$. 
To determine $R$ from $R_x$, a flag $v$ is introduced that indicates whether $R_y$ is greater than $q/2$ or not. That is, $R=(R_x,R_y)$ is uniquely determined by $(R_x,v)$. 
Let $x\xleftarrow{\$}S$ denote that an element $x$ is chosen at random from a set $S$. 

\medskip
\noindent\textbf{ECDSA with key recovery}
\begin{itemize}
\item $\ECDSA.\SigKeyGen(1^\lambda)$: Choose $d\xleftarrow{\$}\mathbb{Z}_q$ and compute $P=dG$. Output $\sigsigk=d$, $\sigvk=P$, and $\addr=\B_{96..255}(H(P))$.  

\smallskip
\item $\ECDSA.\SigSign(\sigsigk,M)$: Choose Choose $r\xleftarrow{\$}\mathbb{Z}_q$, and compute $h=H(M)$ and $R=rG$. Let $R=(R_x,R_y)$. Compute $s=\frac{h+d R_x}{r}\bmod{q}$ and output $\sigma=(s,R_x,v)$. 

\smallskip
\item $\ECDSA.\SigVerify(\addr,\sigma,M)$: Parse $\sigma=(s,R_x,v)$. Recover $R=(R_x,R_y)$ from $(R_x,\allowbreak v)$, and compute $P=\frac{s}{R_x}(R-\frac{h}{s}G)$. Output 1 if $\addr=\B_{96..255}(H(P))$, and 0 otherwise. 
\end{itemize}

\noindent 
From the original verification equation $\frac{h}{s}G+\frac{R_x}{s}P=\frac{h}{s}G+\frac{R_x d}{s}G=\frac{h+R_x d}{s}G=rG=R$, $P=\frac{s}{R_x}(R-\frac{h}{s}G)$ holds. 

Ethereum Yellow Paper~\cite{ETHYellow} stipulates that a signature $\sigma=(s,R_x,v)$ is invalid if $0<s<q/2+1$ does not hold. This is because $(-s,R_x,\bar{v})$ is a valid ECDSA signature when $(s,R_x,v)$ is a valid signature where $\bar{v}$ is the opposite frag of $v$. 
More precisely, if $\frac{h}{s}G+\frac{R_x}{s}P=R$ holds, then $\frac{h}{-s}G+\frac{R_x}{-s}P=-R$ holds. 
The x-coordinate of $R$ and $-R$ are the same and the verification of the original ECDSA checks the x-coordinate only. 
Thus, in the $\ECDSA.\SigSign$, $-s$ is used for generating a signature if $s>q/2+1$. 
We omit this procedure from the description of the $\ECDSA.\SigSign$ algorithm above, for the ease of understanding. 
We do not explicitly consider the range of $s$ anymore in this paper. 

\section{Generic Construction of IBS from Signatures}

In this section, we introduce a generic construction of IBS $\IBS=(\IBSSetup,\allowbreak \IBSKeyDer,\allowbreak \IBSSign,\allowbreak \IBSVerify)$ from a signature scheme $\Sig=(\SigKeyGen,\allowbreak\SigSign,\allowbreak\SigVerify)$~\cite{BellareNN04,DodisKXY03,GentryS02}. 
This construction is so-called \lq\lq Certificate-based Construction" where the KGC generates a certificate $\cert$ for $ID$ using the master secret key $\msk$. In the verification algorithm $\IBSVerify$, the validity of $\cert$ is also checked in addition to the usual signature verification. 

\begin{itemize}
\item $\IBSSetup(1^\lambda)$: The setup algorithm takes a security parameter $\lambda\in\mathbb{N}$. Run $(\mpk,\msk)\leftarrow \SigKeyGen(1^\lambda)$ and output a pair of a master public key and a master secret key $(\mpk,\msk)$. 

\smallskip
\item $\IBSKeyDer(\msk,ID)$: The key derivation algorithm takes $\msk$ and $ID$. Run $(\sigvk,\sigsigk)\allowbreak\leftarrow \SigKeyGen(1^\lambda)$, compute $\cert\leftarrow\SigSign(\msk,\allowbreak \sigvk||ID)$, and output a secret signing key for $ID$ $\ibssigk=(\cert,\sigvk,\allowbreak \sigsigk)$. 

\smallskip
\item $\IBSSign(\ibssigk,M)$: The signing algorithm takes $\ibssigk$ and $M$ to be signed. Parse $\ibssigk=(\cert,\sigvk,\sigsigk)$. Compute $\sigma\leftarrow\SigSign(\sigsigk,\allowbreak M)$ and output a signature $\ibssig=(\sigma,\sigvk,\cert)$. 

\smallskip
\item $\IBSVerify(\mpk,ID,\ibssig,M)$: The verification algorithm takes $\mpk$, $ID$, $\ibssig$, and $M$. Parse $\ibssig=(\sigma,\sigvk,\allowbreak \cert)$. Output 1 if both $\SigVerify(\mpk,\allowbreak \cert,\allowbreak\sigvk||ID)=1$ and $\SigVerify(\sigvk,\sigma,M)=1$ hold, and 0, otherwise.  
\end{itemize}

\noindent
If the underlying signature scheme is unforgeable (i.e., EUF-CMA secure where EUF-CMA stands for existential unforgeability under chosen message attack), then the IBE scheme is also unforgeable (See~\cite{BellareNN04,DodisKXY03,GentryS02}). Basically, no adversary can produce a valid signature under some ID even the adversary obtains signing keys of other identities.

Next, to clarify the case that the above generic construction is instantiated by ECDSA with key recovery, we introduce the IBS scheme as follows. 
Due to the key recovery functionality, we replace $\sigvk$ contained in $\ibssig=(\sigma,\sigvk,\cert)$ to $\addr$. 
Moreover, we replace $\mpk$ that is an input of the $\IBSVerify$ algorithm to $\addr_{\sf KGC}$. 

\begin{itemize}
\item $\IBSSetup(1^\lambda)$: Choose $d_{\sf KGC}\xleftarrow{\$}\mathbb{Z}_q$ and compute $P_{\sf KGC}=d_{\sf KGC}G$. 
Output $\msk=d_{\sf KGC}$, $\mpk=P_{\sf KGC}$, and $\addr_{\sf KGC}=\B_{96..255}(H(P_{\sf KGC}))$. 

\smallskip
\item $\IBSKeyDer(\msk,ID)$: Parse $\msk=d_{\sf KGC}$. Choose $d\xleftarrow{\$}\mathbb{Z}_q$ and compute $P=dG$. 
Choose $r\xleftarrow{\$}\mathbb{Z}_q$ and compute $h_{ID}=H(P||ID)$ and $R=rG$. Let $R=(R_x,R_y)$. Compute $s=\frac{h_{ID}+d R_x}{r}\bmod{q}$ and set $\cert=(s,R_x,v)$. 
Output $\ibssigk=(\cert,P,d)$. 

\smallskip
\item $\IBSSign(\ibssigk,M)$: Parse $\ibssigk=(\cert,P,d)$ and $\cert=(s,R_x,\allowbreak v)$. Let $\addr=\B_{96..255}(H(P))$. 
Choose $r^\prime\xleftarrow{\$}\mathbb{Z}_q$ and compute $h=H(M)$ and $R^\prime=r^\prime G$. Let $R^\prime=(R^\prime_x,R^\prime_y)$. Compute $s^\prime=\frac{h+d R^\prime_x}{r}\bmod{q}$ and set $\sigma=(s^\prime,R^\prime_x,\allowbreak v^\prime)$. 
 Output $\ibssig=(\sigma,\addr,\allowbreak \cert)$. 

\smallskip
\item $\IBSVerify(\addr_{\sf KGC},ID,\ibssig,M)$: Parse $\ibssig=(\sigma,\addr,\allowbreak \cert)$, $\sigma=(s^\prime,R^\prime_x,\allowbreak v^\prime)$, and $\cert=(s,R_x,v)$. Compute $h=H(M)$. 
Compute $R^\prime=(R^\prime_x,R^\prime_y)$ from $(R^\prime_x, v^\prime)$, and compute $P=\frac{s^\prime}{R^\prime_x}(R^\prime-\frac{h}{s^\prime}G)$. Compute $h_{ID}=H(P||ID)$. Compute $R=(R_x,R_y)$ from $(R_x, v)$. Compute $P_{\sf KGC}=\frac{s}{R_x}(R-\frac{h_{ID}}{s}G)$. Output 1 if both $\addr_{\sf KGC}=\B_{96..255}(H(P_{\sf KGC}))$ and $\addr=\B_{96..255}(H(P))$, and 0, otherwise.  
\end{itemize}

\medskip\noindent\textbf{Evaluation on the IBS scheme based on ECDSA with key recovery}. 
Here, we consider whether a vanity address can be generated by using the ECDSA-based IBS scheme via the above attempt. 
First, let us check the impact of introducing $\addr_{\sf KGC}$. Now, all users are required to manage $\cert$ which is a valid ECDSA signature under $\addr_{\sf KGC}$. Here, the corresponding message is $P||ID$, and does not a transaction. 
We may consider a case that a user sets a transaction as $ID$. However, no cryptocurrency is stored on $\addr_{\sf KGC}$ and the corresponding message contains $P$ in addition to $ID$. Thus, we conclude that introducing $\addr_{\sf KGC}$ does not affect the security. 

Second,  let us check the impact of introducing $ID$. Since the underlying ECDSA provides the key recovery functionality, $ID$ needs to be recovered if $ID$ is required to be a verification key. That is, even if we can suitably set $ID$ to produce a vanity address via the procedure (introduced in Our Motivation part), $ID$ additionally needs to satisfy $$ID=\frac{s^\prime}{R^\prime_x}(R^\prime-\frac{h}{s^\prime}G)$$ 
This indicates that the KGC needs to find $d$ satisfying $ID=dG$, and it requires the same procedure of the classical try-and-error method. Moreover, $\mpk$ is also required for signature verification (i.e., even if the hash value of $ID$ can be set as a desired vanity address, signatures are not verified by $ID$ only). 

To sum up, the IBS scheme is not directly employed to generate a vanity address because: 
\begin{itemize}
\item Due to the verification process of ECDSA with key recovery, $ID$ needs to be recovered from a ECDSA signature and a message in the ECDSA-based IBS scheme that requires the same procedure of the classical try-and-error method. 

\item Due to the syntax of IBS where, in addition to $ID$, the master public key is required for running the verification algorithm. Here, the master public key is an ECDSA verification key in the ECDSA-based IBS scheme. 
\end{itemize}

\section{Proposed System}
\label{proposed}

Due to the evaluation in the previous section, we assign a desired value to $ID$ directly (this is not the same as the procedure introduced in Our Motivation part). Let $\addr=\B_{96..255}(H(P))$ be an address of a user where $P$ is contained in $\ibssigk=(\cert,P,d)$. If $P$ is connected to $ID$ (i,e, $\cert$ is a valid ECDSA signature on $P||ID$), we say that the user of $\addr=\B_{96..255}\allowbreak(H(P))$ is $ID$. 

\begin{itemize}
    \item A user selects $ID$ (as a desired value such as the user's name), and obtains $\ibssig=(\sigma,\addr,\allowbreak \cert)$ from the KGC. The user sets $\addr$ as own address and opens $\addr$ together with $ID$. 

    \smallskip
    \item When the user issues a transaction $M$, the user generates $\ibssig=(\sigma,\addr,\allowbreak \cert)$ using the 
    $\IBSSign$ algorithm. Set $(ID,\sigma,\cert)$ be a signature on $M$. 

    \smallskip
    \item A transaction verifier recovers $P$ from $\sigma=(s^\prime,R^\prime_x,v^\prime)$ and also recovers $P_{\sf KGC}$ from $P$, $ID$, and $\cert=(s,R_x,v)$. The verifier accepts that the transaction issuer is $ID$ if both $\addr_{\sf KGC}=\B_{96..255}(H(P_{\sf KGC}))$ and $\addr=\B_{96..255}(H(P))$.  
\end{itemize}

\noindent
Since two signatures are verified by the transaction verifier, introducing the IBE scheme does not affect security compared to the case that ECDSA with key recovery is employed. 
More concretely, the procedure that the transaction verifier recovers $P$ from $\sigma=(s^\prime,R^\prime_x,v^\prime)$ and checks $\addr=\B_{96..255}(H(P))$ is the totally the same as the transaction verification procedure in Ethereum. Additionally, the transaction verifier recovers  $P_{\sf KGC}$ from $P$, $ID$, and $\cert=(s,R_x,v)$ and checks $\addr_{\sf KGC}=\B_{96..255}(H(P_{\sf KGC}))$. We remark that the verification of $\cert$ (i.e., verification of whether the user of $\addr$ is $ID$) is independent to the underlying transaction $M$. Thus, $\cert$ only needs to be verified once, and introducing the IBE scheme does not affect security in terms of transaction verification.

Note that we need to consider the key escrow problem where the KGC also can generate signatures. 
To solve the problem, TEE (Trusted Execution Environment) could be employed: 
Assume that an enclave stores the master secret key. A user sends $ID$ and a public key ${\sf pk}$ of a public key encryption scheme to a TEE via a remote attestation. Then, the TEE generates a secret singing key for $ID$ on the enclave, encrypts the secret signing key by ${\sf pk}$, and returns the ciphertext to the user. Then, the user can obtain the secret signing key by decrypting the ciphertext. Our work is regarded as the first stepping stone to employ IBS in the blockchain environment and further evaluation of the key generation procedure is left as a future work of this paper.

\section{Efficiency Evaluation}

In this section, we implement the IBS scheme based on ECDSA with key recovery by Solidity, and check the gas cost for the signature verification. To the best of our knowledge, no other scheme employing IBS for generating vanity address has been considered, as mentioned in Related Work section. Moreover, no attempt to construct an IBE scheme from ECDSA with key recovery also exists. In this perspective, we compare the performance of our system with ECDSA with key recovery. 
Intuitively, the gas cost is twice compared to that of the ECDSA signature verification since two ECDSA verification procedures are run in the IBS signature verification procedure, for $\cert$ and $\sigma$. 
As mentioned above, the verification of $\cert$ is independent to the underlying transaction $M$ and $\cert$ only needs to be verified once.  That is, the gas cost is almost same as that of the ECDSA signature verification after $\cert$ has been verified. 

Our implementation environment is described as follows: Precision Tower3431 (Processor: 3.10 Ghz octa-core Intel Core i9, Memory: 16 GB). For signature generation, we employed libraries, ethereumjs-util and ethereumjs-wallet, which are run on node.js (v20.17.0). For signature verification, we implement a smart contract using Solidity (we indicates the version as $>=$0.7.0$<$0.9.0). 
We set $ID$ as 128-bit value since it can express 13 characters by ASCII codes and seems sufficient to express a name, an e-mail address, and so on.

Our Solidity code is described in Listing~\ref{IBSverify} as follows. 
Here, let MSG be a message to be sent, SIGNER\_ADDRESS be the address of the transaction issuer, (s, Rx, v) be a signature on MSG, SIG\_PBK\_ID be a strong that contains the issuer's public key and arbitrary string ($P||ID$), KGC\_ADDRESS be the address of the KGC ($\addr_{\sf KGC}$), and (CERT\_s, CERT\_Rx, CERT\_v) be a signature on SIG\_PBK\_ID. 
Let the function ECDSA.Sig.Verify() be a verification algorithm that checks whether MSG is sent from the issuer and the function Cert.Verify() be a verification algorithm that checks whether the public key of the transaction issuer is associated to $ID$. 

\begin{lstlisting}[caption=Example of basic Solidity Code of IBS Signature Verification,label=IBSverify]
1  string MSG;
2  address SIGNER_ADDRESS;
3  bytes32 s;
4  bytes32 Rx;
5  uint8 v;
6
7  string SIG_PBK_ID;
8  address KGC_ADDRESS;
9  bytes32 CERT_s;  
10 bytes32 CERT_Rx;
11 uint8 CERT_v;
12 
13 function ECDSA.Sig.Verify() public view returns (bool) {
14   bytes32 msgHash = keccak256(bytes(MSG));
15   address signer = ecrecover(msgHash, s, Rx, v);
16   if (signer == SIGNER_ADDRESS) {
17       return true;
18   } else {
19       return false;
20   }
21 }
22
23 function Cert.Verify() public  view returns (bool){
24  bytes32 msgHash = keccak256(bytes(SIG_PBK_ID));
25  address signer = ecrecover(msgHash, CERT_s, CERT_Rx, CERT_v);
26  if (signer == KGC_ADDRESS) {
27       return true;
28  } else {
29      return false;
30  }
31}
\end{lstlisting}

We denote gas costs for running each function in Table~\ref{table:gas}. We used USD price on October 22, 2024. 
Remark that $\IBSVerify(\addr_{\sf KGC},ID,\ibssig,M)$ runs both ECDSA.Sig.Verify() and Cert.Verify(). 

\begin{table}[h]
  \caption{Gas Costs}
  \label{table:gas}
  \centering
  \begin{tabular}{|c|r|r|}
    \hline
     & ECDSA.Sig.Verify()  & Cert.Verify() \\
    \hline \hline
    Gas & 21,849  &26,251 \\
    USD & 0.77 & 0.92 \\
    \hline
  \end{tabular}
\end{table}

We expected that the verification costs of IBS signature constructed by ECDSA with key recovery is twice as those of ECDSA with key recovery. However, the actual gas cost is approximately 2.2 times higher. The reason behind is that the cost of Cert.Verify() is 1.2 times higher than that of ECDSA.Sig.Verify(). Here, the message to be signed is $P||ID$ in Cert.Verify() and the size of $P$ is 64 bytes (512 bits) (See Ethereum Yellow Paper~\cite{ETHYellow}) and the size of the message is 640 bits. On the other hands, the message $M$ is a transaction encoded by RLP (Recursive Length Prefix) and is represented by 256 bits. Since the hash value of the message is computed in the ECDSA signature verification procedure, the difference of the underlying message size is the main reason why Cert.Verify() requires a higher gas cost than that of ECDSA.Sig.Verify(). 
As above mentioned, however, the verification of $\cert$ is independent to the underlying transaction $M$ and $\cert$ only needs to be verified once. That is, the gas cost for verifying a transaction is almost same as that of the ECDSA signature verification after $\cert$ has been verified. 

\section{Conclusion}

In this paper, we consider whether IBS can be employed to generate a vanity address and demonstrate that the IBS scheme (constructed by ECDSA with key recovery) is not directly employed to generate a vanity address. 
As the next attempt, we propose a method to connect any value to an address using IBS. 
The actual gas cost for verifying a transaction is almost same as that of the ECDSA signature verification after $\cert$ has been verified. 
Since the experimental simulation is overly simple, considering comprehensively blockchain performance is left as a future work. 
Moreover, implementation evaluation of the key generation procedure using TEE is also an important future work. 

We remark that we do not deny any possibility to generate a vanity address safety and efficiently via an IBS scheme. Even if we turn blind eye the key recovery functionality of ECDSA, we need to consider how to treat $\mpk$ when an IBS signature is verified. We may be able to employ the Barreto et al. IBS scheme~\cite{BarretoLMQ05} because Liu et al.~\cite{LiuYWNW19} proposed a signature scheme (for enhancing the security of stealth address) using an IBE scheme that does not require $\mpk$ for the signature verification procedure, and introduced that the Barreto et al. IBS scheme as such an IBS scheme. Further considering vanity address generation methods via IBS is left as a future work. 

\medskip
\noindent\textbf{Acknowledgment}: The authors thank Mr. Kota Chin for his invaluable comment against Ethereum vanity address. 
This work was supported by JSPS KAKENHI Grant Numbers JP21K11897, JP23K24844, and JP25H01106.

\end{document}